\documentclass[%
 reprint,
 floatfix, amsmath,amssymb,
 aps,
]{revtex4-1}

\usepackage{graphicx}
\usepackage{dcolumn}
\usepackage{bm}
\usepackage{pythontex}
\usepackage{amsmath}
\usepackage{hyperref}
\usepackage[short]{optidef} 
\usepackage[version-1-compatibility]{siunitx}

\usepackage[usenames,dvipsnames]{xcolor}

\newcommand{\ket}[1]{|#1\rangle}

\begin{document}

\title{Benchmarking Quantum Annealing Controls with Portfolio Optimization}

\author{Erica Grant}
 \altaffiliation[Also at ]{Bredesen Center,\\
 University of Tennessee\\
 Knoxville, TN, 37996}
 \email{egrant8@vols.utk.edu}
\author{Travis S.~Humble}%
 \altaffiliation[Also at ]{Bredesen Center,\\
 University of Tennessee\\
 Knoxville, TN, 37996}
 \email{humblets@ornl.gov}
\affiliation{%
 Quantum Computing Institute, Oak Ridge National Laboratory \\
 Oak Ridge, TN, 37830
}%

\author{Benjamin Stump}
 \email{stumpbc@ornl.gov}
\affiliation{
 National Transportation Research Center\\
 Oak Ridge National Laboratory\\
 Knoxville, TN, 37932
}%

\begin{abstract}
Quantum annealing offers a novel approach to finding the optimal solutions for a variety of computational problems, where the quantum annealing controls influence the observed performance and error mechanisms by tuning the underlying quantum dynamics. However, the influence of the available controls is often poorly understood, and methods for evaluating the effects of these controls are necessary to tune quantum computational performance. Here we use portfolio optimization as a case study by which to benchmark quantum annealing controls and their relative effects on computational accuracy. We compare empirical results from the D-Wave 2000Q quantum annealer to the computational ground truth for a variety of portfolio optimization instances. We evaluate both forward and reverse annealing methods and we identify control variations that yield optimal performance in terms of probability of success and probability of chain breaks.
\end{abstract}
\maketitle

\section{\label{sec:Introduction} Introduction}
Optimization is integral to many scientific and industrial applications of applied mathematics including verification and validation, operations research, data analytics, and logistics, among others \cite{pardalos1987constrained, tsai2014optimization}. In many cases, exact methods of solution, including stochastic optimization and quadratic programming, are computationally intractable and novel heuristics are used frequently to solve problems in practice \cite{krentel1986complexity}. Quantum annealing (QA) offers a novel meta-heuristic that uses quantum mechanics for unconstrained optimization by encoding the problem cost function in a Hamiltonian \cite{farhi2000quantum,morita2008mathematical}. Recovery of the Hamiltonian ground state solves the original optimization problem and this approach has been mapped to a variety of application areas \cite{djidjev2018efficient, neukart2017traffic, stollenwerk2019quantum, martovnak2004quantum}. Several experimental efforts have realized  quantum annealers \cite{johnson2011quantum, lanting2014entanglement, van_der_Ploeg_2007}, and application benchmarking of these systems has shown QA is capable of finding the correct result with varying probability of success \cite{katzgraber2014glassy, king2015benchmarking,zhu2016best,jarret2016adiabatic,o2018nonnegative,albash2018demonstration,ajagekar2020quantum}.
\par 
QA performance depends implicitly on the complexity of the underlying problem instance as well as the controls that implement the heuristic \cite{venturelli2019reverse,quiroz2019robust}. Presently, there are multiple controls available to program quantum annealers that may each impact the observed probability of success. Notionally, the controls may be categorized as pre-processing, annealing, and post-processing methods. Whereas pre-processing controls define the encoded Hamiltonian and embedding onto the quantum annealer \cite{vinci2015quantum, bian2016mapping}, the annealing controls drive the time-dependent physics of the device and the underlying quantum state \cite{marshall2019power, venturelli2019reverse} while post-processing controls influence the read-out and decoding of the observed results \cite{pudenz2014error, pudenz2016parameter}. Collectively, the choice for each type of control may either enhance or impede the probability of reaching the encoded ground state and, therefore, impact the resulting solution state. 
\par
Here we benchmark a selection of pre-processing and annealing controls available in a programmable quantum annealer \cite{johnson2011quantum} using a well-defined class of unconstrained optimization problems derived from the application of Markowitz portfolio theory \cite{markowitz1952portfolio}. As a variant of binary optimization, Markowitz portfolio optimization selects the subset of investment assets expected to yield the highest return value and minimal risk while staying within a total budget constraint \cite{markowitz1952portfolio, elsokkary2017financial}. We cast this problem which forms a complete graph as unconstrained optimization and benchmark the probability of success for QA to recover the global optimum. In particular, we benchmark the pre-processing and annealing controls available in the 2000Q, a programmable quantum annealer from D-Wave Systems \cite{johnson2011quantum}. This includes controls for mapping the logical problem onto hardware and scheduling the annealing process. We gather insight into the underlying dynamics using multiple measures of success tested across an ensemble of randomly generated instances of portfolio optimization.
\par 
Previous research has benchmarked QA in comparison to classical heuristics for solving various optimization problems \cite{mcgeoch2013experimental, king2015benchmarking, steiger2015heavy}. In particular, several variations of portfolio optimization have been used  to benchmark QA performance \cite{marzec2016portfolio, venturelli2019reverse, rosenberg2016solving}. Rosenberg et al.~demonstrated several encodings of a multi-period Markowitz portfolio optimization formulation to be solvable using a quantum annealer and found promising initial results in probability to find the optimal result \cite{rosenberg2016solving}. Venturelli et al.~benchmarked a similar mean-variance model of portfolio optimization using a hybrid solver that couples quantum annealing with a genetic algorithm \cite{venturelli2019reverse}. This hybrid algorithm was found to be 100x faster than forward annealing alone. In this work, we present a formulation of portfolio optimization to benchmark the behaviour of QA controls. We present studies focused on the variability in success with respect to available quantum annealing controls in an attempt to establish a methodology for finding an optimal set of controls which yield the highest solution quality \cite{pelofske2019optimizing, king2014algorithm}.
\par
The presentation is organized as follows. In Sec.~\ref{sec:Quantum Annealing}, we review quantum annealing and the the available controls. In Sec.~\ref{sec:Methods}, we provide an overview of the benchmarking methods and the use of Markowitz portfolio selection for problem specification. In Sec.~\ref{sec:Results}, we present the results from experimental testing using different quantum annealing controls with the 2000Q. We offer conclusions in Sec.~\ref{sec:level1}. 
\section{Quantum Annealing \label{sec:Quantum Annealing}}
Under ideal conditions, forward annealing evolves a quantum state $\ket{\Psi(t)}$ under the time-dependent Schr\"{o}dinger equation 
\begin{equation}
\label{Schrodinger Equation}
i \hbar \frac{\partial}{\partial t} \ket{\Psi(t)}  = H(t) \ket{\Psi (t)} \hspace{1cm} t \in [0, T]
\end{equation}
where $T$ is the total forward annealing time and the time-dependent Hamiltonian is
\begin{equation}
\label{Adiabatic Evolution}
H(t) = A(s(t))H_{0} + B(s(t)) H_{1}.
\end{equation}
where $s(t) \in [0,1]$ is the control schedule and time-dependent amplitudes $A(s)$ and $B(s)$ satisfy the conditions $A(0)\gg B(0)$ and  $A(1) \ll B(1)$. We consider the initial Hamiltonian $H_0 = - \sum_i^n \sigma_i^x$ as a sum of Pauli-$X$ operators $\sigma_i^x$ over $n$ spins. The final Hamiltonian $H_1$ represents the unconstrained optimization problem with a corresponding ground state that encodes the computational solution. We will consider below only problems represented using the Ising Hamiltonian
\begin{equation}
\begin{aligned}
\label{eq:Ising_Hamiltonian}
H_1 = \sum_{i} h_i \sigma_{i}^z + \sum_{i,j} J_{i,j} \sigma_{i}^z \sigma_{j}^z + \beta
\end{aligned}
\end{equation}
where $h_i$ is the bias on the $i^{th}$ spin, $J_{i,j}$ is the coupling strength between the $i^{th}$ and $j^{th}$ spin, $\sigma_{i}^z$ is the Pauli-$Z$ operator for the $i^{th}$ spin, and $\beta$ is a problem-specific constant. The Ising Hamiltonian is well known for representing a variety of unconstrained optimization problems \cite{lucas2014ising}.
\par 
Instantaneous eigenstates at time $t$ are defined as
\begin{equation}
\label{instaneous_eigenstates}
H(t) \ket{\Phi_j(t)} = E_j (t) \ket{\Phi_j(t)}
\end{equation}
where $j$ ranges from $0$ to $N-1$ with $N=2^n$ the dimension of the Hilbert space.  
For an initial quantum state prepared in the lowest-energy eigenstate at time $t=0$, i.e, $\ket{\Psi(0)} = \ket{\Phi_0(0)}$, adiabatic evolution under the Hamiltonian in Eq.~(\ref{Adiabatic Evolution}) to time $T$ will prepare the final state $\ket{\Psi(T)} = \ket{\Phi_0(T)}$ with high probability provided $T$ is sufficiently large. In particular, the evolution must be much longer than the inverse square of the minimum energy gap between the ground and first excited states \cite{farhi2000quantum}. At time $T$, the prepared quantum state is measured in the computational basis to generate a candidate solution for the encoded problem.
\par
Another variation of quantum annealing reverses the time-evolution process by beginning in an eigenstate of $H_1$. Known as reverse annealing, the initial quantum state evolves under Eq.~(\ref{Adiabatic Evolution}) in the reverse direction. The Hamiltonian starts as $H_1$ at time $t = 0$ and evolves backward to a point $s_{p}$ in the control schedule that corresponds to time $t_{1}$. The Hamiltonian then pauses for a time $t_p = t_{2} - t_{1}$ before evolving in the forward direction from the value $s_{p}$ at time $t_{2}$ back to the final Hamiltonian at time $T'$, where the latter time represent the reverse annealing time. The control schedule for reverse annealing is then defined as \cite{Yamashiro_2019, Passarelli_2020}
\begin{equation}
    s'(t)= \begin{cases} 
        1 + \frac{(s_{p} - 1)}{t_1}t, & \ 0 \leq t \leq t_{1}\\ 
        s_{p}, & t_{1} \leq t \leq t_{2}\\
        s_{p} + \frac{(1 - s_{p})}{(T' - t_{2})} (t - t_{2}) & t_{2} \leq t \leq T'
        \end{cases}
\label{eq: RA s(t)}
\end{equation}
\par
\begin{figure}[h!]
\centering
\includegraphics[width=85mm]{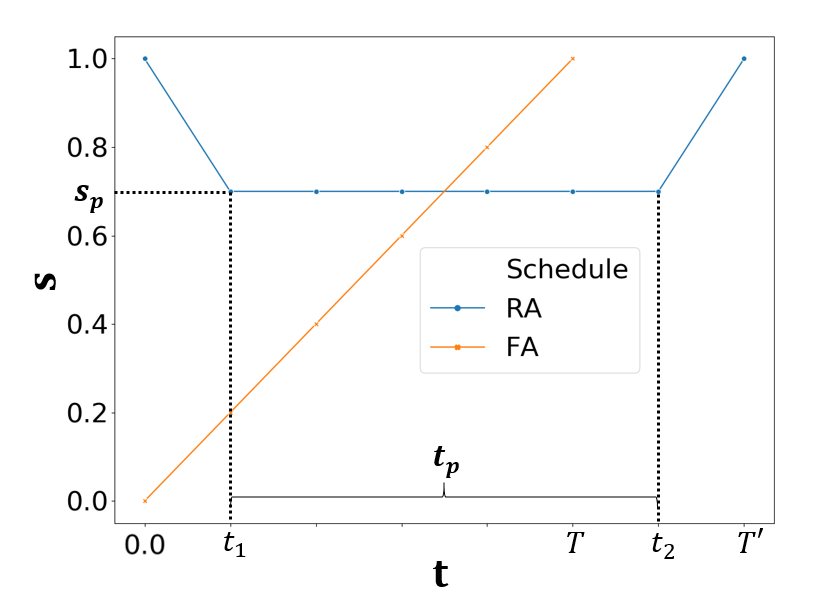}
\caption{The control schedule for reverse annealing (RA) compared to forward annealing (FA) plotted with respect to time. The control schedule for forward annealing starts at $t = 0, s = 0$ and anneals at a constant rate to $t = T, s = 1$, while the control schedule for reverse annealing starts at $t = 0$ with $s = 1$, decreases to a value $s_{p}$ at time $t_1$, pauses for time $t_p = t_2 - t_1$, and then increases to $s = 1$ at time $T'$.}
\label{fig:reverse_annealing}
\end{figure}

 
The differences in the control schedules of forward and reverse annealing are demonstrated in Fig.~\ref{fig:reverse_annealing}, where a linear reverse annealing schedule is compared to a linear forward annealing schedule using the amplitudes $A(s) = (1 - s)$ and $B(s) = s$. Notably, forward annealing controls increase monotonically with time whereas reverse annealing controls include a change in the direction of the control schedule where the ramp time from $s = 1$ to $s_p$ is $t_r = t_{1}$, the time paused at $s_p$ is $t_p$, and the quench time back from $s_p$ to $s = 1$ is $t_q = T^\prime - t_{2}$. 
\subsection{Quantum Annealing Controls \label{sec:Controls}}
In practice, there are non-ideal behaviours that arise in practical implementations of quantum annealing. Equations (\ref{Schrodinger Equation})-(\ref{eq: RA s(t)}) represent quantum annealing under ideal adiabatic conditions that are difficult to realize in actual quantum devices. Real-world quantum annealers have limits in the ability to control the Hamiltonian and quantum dynamics \cite{pearson2019analog}. In addition, the presence of ill-characterized environmental couplings give rise to flux noise \cite{martinis2003decoherence}. The imperfect setting of the Hamiltonian parameters $(h, J_{i,j})$ by the analog control circuits gives rise to a small intrinsic control error \cite{king2014algorithm}. These errors undermine the accuracy of the physical hardware \cite{vinci2015quantum, pearson2019analog}. Finally, annealing too quickly may violate the essential adiabatic condition \cite{farhi2000quantum}, while annealing too slowly may lead to undesired thermal excitations of the quantum state due non-zero temperature fluctuations \cite{novikov2018exploring}. This multitude of effects complicates both the description of quantum annealing as well as the assessment of its performance.
\par 
Given the implicit dependence on several competing factors, a variety of strategies have emerged for controlling quantum annealing to maximize probability of success in recovering the ground state and minimizing errors in the quantum computational solution. These control strategies include  efficiently mapping the problem Hamiltonian onto the physical hardware Hamiltonian, tuning annealing schedule, applying variable transformations to mitigate control biases, and using reverse annealing to refine initial solutions \cite{king2014algorithm, Yamashiro_2019}. 
\par 
We investigate a subset of controls available in the D-Wave 2000Q, a programmable quantum annealer composed from an array of superconducting flux qubits operated at cryogenic temperatures \cite{Bunyk_2014}. The 2000Q consists of up to $2048$ physical qubits arranged in a sparsely connected array  whose governing Hamiltonian is described by a time-dependent, transverse Ising Hamiltonian \cite{tichy2017quantum} for which with the Hamiltonian parameters in the device can be programmed individually. This enables a broad variety of computational problems, including portfolio optimization, to be realized. We briefly review some of the controls available to influence the success of solving these problems using quantum annealing.
\subsubsection{Problem Embedding \label{sec:Embedding}}
\begin{figure}[h!]
\centering
\includegraphics[width=60mm]{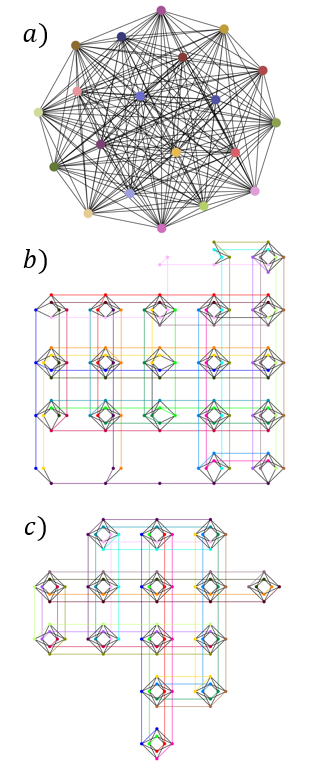}
\caption{The embedding of a $20$ logical spin complete graph onto a Chimera graph structure. Figure $a)$ is complete $K_{20}$ graph which is fully connected with $20$ nodes and $190$ edges where each node represents a logical spin and each edge is a coupling between spins. Figure $b)$ is the CMR algorithm which requires the allocation of $23$ unit cells and $c)$ is the clique embedding algorithm which requires the allocation of $15$ unit cells. The nodes represent physical qubits, lines are the couplings between physical qubits, and each color is a different physical spin chain corresponding to a logic spin. } 
\label{fig:embeddings}
\end{figure}

The Hamiltonian encoding the computational problem must be mapped into the physical hardware while satisfying the constraints of limited connectivity. The 2000Q hardware supports a sparse Chimera graph in which physical qubits are not fully connected but have average degree 6. A fully connected graph, like in Fig.~\ref{fig:embeddings}, must be mapped onto the more sparse Chimera graph. A single spin from the input Hamiltonian may be realized in hardware using multiple physical qubits that form a strongly interacting representative chain of spins. By judiciously choosing these chains and their interactions, the original input Hamiltonian may be constructed. This process, known as embedding, depends on the input problem as well as the target hardware connectivity. In general, embedding is NP-hard for arbitrary input graphs \cite{choi2008minorembedding}, and there are upper limits on the maximum graph that can be embedded \cite{klymko2012adiabatic}. For example, the largest fully connected problem that can be embedded onto the 2000Q has $\sim 60$ spins, while the limit in practice depends on the number of faulty/inactive physical qubits in the device. 
\par
Embedding algorithms that optimize chain length may greatly reduce the number of physical qubits required by considering problem symmetry as well as the location of faults in the hardware. We highlight two embedding algorithms widely used in programming the 2000Q. The first method by Cai, Macready, and Roy is based on randomized placement and search for the weighted shortest path between spin chains \cite{cai2014practical}. This method, which we denote as CMR, applies to arbitrary input graphs but typically creates a distribution of chain lengths. By contrast, a second method by Boothby, King, and Roy based on a clique embedding typically generates shorter and uniform chain lengths of size 
\begin{equation}
    l_c = \frac{n}{4} + 1
\label{eq:clique_chain_length}
\end{equation}
for $n$ logical spins \cite{boothby2015fast}.  A representative example of the output from these different methods is shown in Fig.~\ref{fig:embeddings} using a fully connected problem with $20$ logical spins. Both methods are available in the D-Wave Ocean software library \cite{embedding_tools}. 
\par
Ensuring an embedded chain of qubits collectively represents a single logical variable requires an intra-chain coupling that is larger in magnitude than the the inter-chain couplings between chains. In other words, the chain of physical qubits must be strongly coupled to remain a single logical spin. However, it is possible for chains to become ``broken'' in so far as individual physical spins within the chain differ in their final state. In general, chain breaks arise from non-adiabatic dynamics that lead to local excitation out of the lowest energy state with longer chains more susceptible to these effects \cite{king2014algorithm, Dziarmaga_2005}.
\par 
An additional control is required for decoding embedded chains to recover the computed logical spin state. In the absence of chain breaks, the logical value is inferred directly from the unanimous selection of a single spin state by every physical qubit. In the presence of chain breaks, several strategies may be employed to decide the logical value including majority vote \cite{king2014algorithm}, which selects the logical spin value as the value that occurs with the highest frequency in a chain.
\subsubsection{Spin Reversal \label{sec:Control_SpinReversal}}
Interactions between embedded chains arise from the required coupling between the logical spins. However, imperfections in the control of these spins lead to small biases that can become non-negligible for larger qubit chains and contribute to the complex dynamics describing the device. In turn, the probability for finding the expected ground state solution can decrease do to these bias errors. The influence of these errors on the computational result may be mitigated by using spin reversal transforms to average out biases.
\par 
As a gauge transformation, spin reversal redefines the Hamiltonian by replacing the biases and couplings for a subset of spins with their negated value \cite{king2014algorithm, pelofske2019optimizing}. This transformation maintains the ground state of the logical problem. However, this transformation flips the sign of randomly selected qubits so that on average their bias is reduced. This strategy mitigates errors on individual spins by balancing the noise on the device prior to annealing \cite{pudenz2016parameter}. The number of selected spins as well as the parameter $g$ that defines the number of times to perform the transformation. 
\subsubsection{Annealing Schedules \label{sec:Control_AnnealTime}}
\begin{figure}[h!]
\centering
\includegraphics[width=85mm]{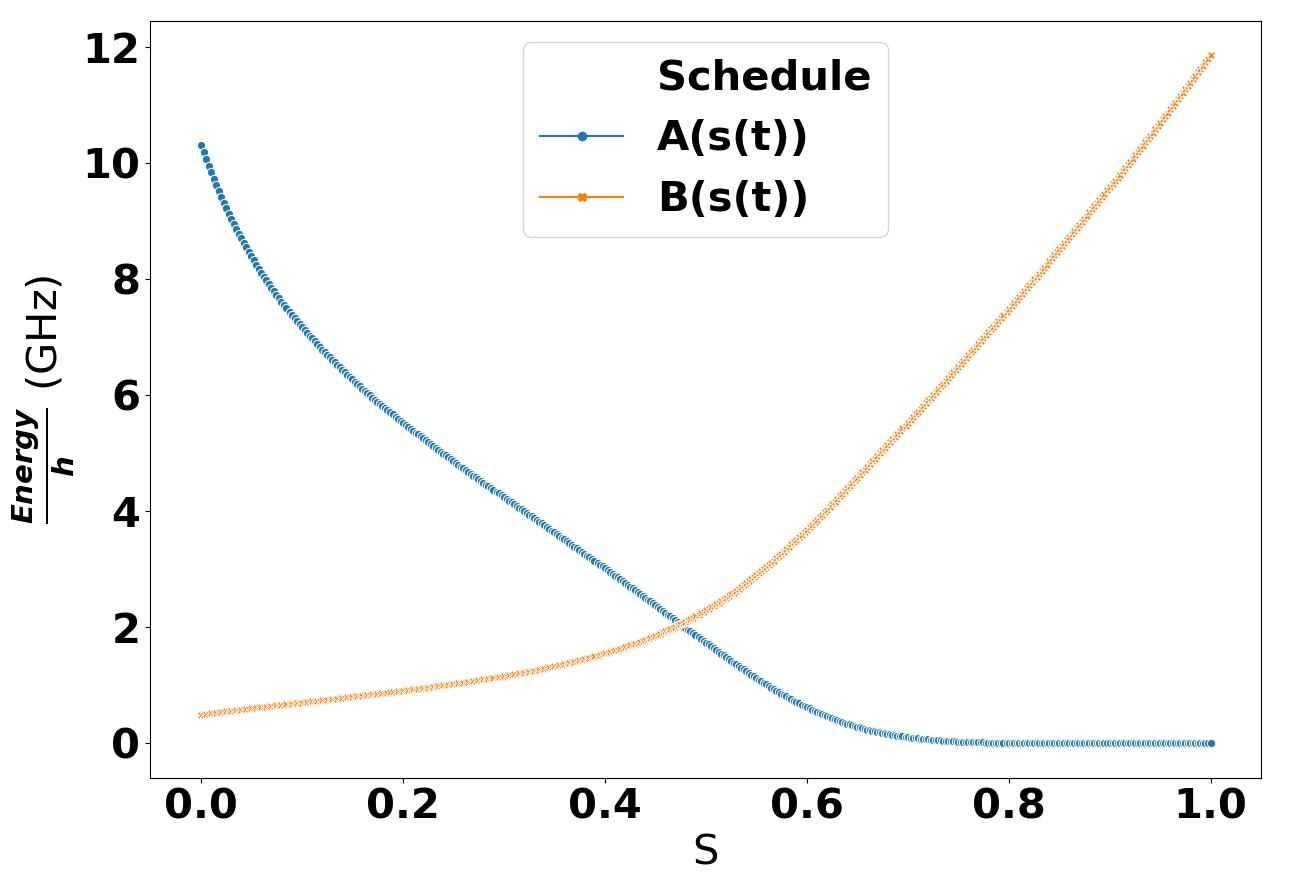}
\caption{The amplitudes of the D-Wave 2000Q over the range of control schedule as measured from $s = 0$ to $s = 1$ in increments of $0.001$.} 
\label{fig:actual_schedules}
\end{figure}

Tailoring the annealing amplitudes $A(s)$ and $B(s)$ is perhaps the most direct method to control forward annealing. The annealing schedules control the rate of change of the $H(t)$, which must be sufficiently slow to approximate the adiabatic condition \cite{childs2001robustness}. An example of the amplitudes in a D-Wave 2000Q is shown in Fig.~\ref{fig:actual_schedules}. While forward annealing on the D-Wave 2000Q, $A(s(t)) >> B(s(t))$ at $t = 0$, $A(s(t))$ decreases and $B(s(t))$ increases for $0 < t < T$, and $B(s(t)) >> A(s(t))$ at $t = T$. 
\par 
The optimal annealing time is problem dependent and inversely proportional to the minimum energy gap \cite{farhi2000quantum}, and, in general, the value and position of the minimum energy gap for a given $H(t)$ is typically unknown and hard to identify. Extending the annealing time $T$ arbitrarily long may not only be limited by hardware parameters but also be counter-productive due to competing thermal processes that depopulate the ground state \cite{pudenz2014error, albash2015decoherence}. There is an upper limit to the total job time $(N_s  T \leq$ 1 s) as well as total annealing time $(T \leq 2$ s) on the D-Wave 2000Q.
\par 
When reverse annealing, the three primary parameters for control are the initial state $e_{i}$, the pause point $s_p$, and the pause duration $t_p$. The times $t_r$ and $t_q$ can also be manipulated, but we keep these constant and symmetric for our experiments. Reverse annealing uses  $e_{i}$ to set the initial quantum state, which may be based on the output of forward annealing, a heuristically computed candidate, a random state or other methods. Our experiments use a pre-computed initial state, e.g., using forward annealing. An iterative procedure is then used which replaces the $e_i$ of each subsequent reverse annealing sample with the output from previous reverse annealing iteration. 
\subsection{Quantum Annealing Metrics}
We characterize quantum annealing using the probability of success
\begin{equation}
    p_{s} = |\langle\Phi_{0}(T)|\rho|\Phi_{0}(T)\rangle|^2
\label{eq: p_s intro}
\end{equation}
defined as the overlap of the final, potentially mixed quantum state $\rho$ prepared by QA with the pure-state describing the expected computational outcome $\Phi_{0}(T)$. Empirically, the probability of success is estimated from the frequency with which the observed solution state matches the expected outcome. When the expected ground state solution is known, we define the statistic $\delta_{i} = 1$ if the $i$-th sample matches the known ground state and $\delta_{i} = 0$ if it does not. For the $k$-th problem Hamiltonian instance, the estimated probability of success is then defined as
\begin{equation}
\tilde{p}_{s}^{(k)} = \frac{1}{N_s}\sum_{i=1}^{N_s}{\delta_{i}}
\end{equation}
where $N_s$ is the total number of samples. The average over an ensemble of $N_{p}$ problem instances is defined as 
\begin{equation}
\tilde{p}_{s} = \frac{1}{N_p}\sum_{k}^{N_p}{ \tilde{p}_{s}^{(k)}}.
\end{equation}
\par
A second metric for characterizing quantum annealing performance, and especially the non-adiabatic dynamics, is the number of chain breaks observed in the recovered solution samples. As noted above, a chain break is observed when the chain of physical qubits embedding a logical spin has more than one unique spin value. We estimate the probability of chain breaks for a problem instance 
\begin{equation}
\begin{aligned}
\tilde{p}_{b}^{(k)} = \frac{1}{N_s}\sum_{i=1}^{N_s}{\epsilon_{i}}
\end{aligned}
\end{equation}
where the statistic $\epsilon_{i} = 1$ when the $i$-th sample solution contains at least one broken chain for any of the logical spins and $\epsilon_{i} = 0$ when no embedded chain is broken. The average probability of chain breaks over an ensemble of $N_{p}$ problem instances is then defined as 
\begin{equation}
\begin{aligned}
\tilde{p}_{b} = \frac{1}{N_p}\sum_{k}^{N_p}{ \tilde{p}_{b}^{(k)}}.
\end{aligned}
\end{equation}
It is important to note that the effects of chain breaks can be mitigated by post-processing methods, such as majority vote, which make hard decisions on the logical spin value.
\par
While the above metrics quantify the probability with which quantum annealing recovers the correct solution, additional information about computational performance comes from the distribution of all solution samples obtained. In particular, the distribution over sample energies provides a representation for the weight of errors in the solution samples. A distribution concentrated around the lowest energy indicates a small number of errors in the computed solutions, while a broad or shifted distribution hints at a larger number of errors.  We denote the energy computed from the $i$-th solution sample as $E(i)$ and we define the $j$-th energy bin as $h_j$. The bin $h_j$ counts the number of samples with an energy in the range $[j, j+1]\Delta$ where $\Delta$ controls the granularity of binning the energies. The resulting set $\{(j\Delta, h_{j})\}$ approximates the energy distribution of the sampled solutions.
\section{\label{sec:Methods} Benchmarking Methods}
We benchmark performance of the quantum annealing controls presented in Sec.~\ref{sec:Quantum Annealing} using a variant of constrained optimization derived from Markowitz portfolio theory. We recast this problem as unconstrained optimization before reducing to quadratic unconstrained binary optimization (QUBO) form. The latter form is easily translated to the classical Ising spin Hamiltonian and, subsequently, to the problem Hamiltonian defined by Eq.~(\ref{eq:Ising_Hamiltonian}).
\subsection{\label{sec:Markowitz}Markowitz Portfolio Selection}
Portfolio optimization selects the best allocation of assets to maximize expected returns while staying within the budget and minimizing  financial risk. The Markowitz theory for portfolio selection focuses on diversification of the portfolio for risk mitigation \cite{markowitz1952portfolio}. Instead of allocating high percentages of a budget toward assets with the highest projected returns, the budget is distributed over assets that minimize correlation between the asset's historical prices. In this model, the covariance between purchasing prices serves as a proxy for risk in which positively correlated assets are considered to be more risky. We review the methods by which the benchmark problems are generated and solved in this section.
\par
We consider Markowitz portfolio optimization as a quadratic programming problem that determines the fraction of available budget $b$ to allocate toward purchasing assets with the goal of maximizing returns while minimizing risk. By selecting a partition number $w$, the fraction $p_w = \frac{1}{2^{(w -1)}}$ represents the granularity of the partition. The portfolio optimization problem selects how many of those partitions to allocate toward each asset with an integer $z_u$. Thus, the fraction of $b$ to invest in each $u^{th}$ asset is given by $p_w b z_u$, and portfolio optimization identifies how much of the $m$ assets to select given the budget $b$ and a risk threshold $c$. Thus, portfolio selection is cast as 
\begin{maxi}
  {z}{\sum_{u = 1}^{m}  r_{u} z_{u}}{}{}
  \addConstraint{\sum_{u = 1}^{m} p_w b z_{u} = b} 
  \addConstraint{\sum_{u,v = 1}^{m} c_{u, v} z_u z_v \leq c}
  \label{eq:MPO_classic}
 \end{maxi}
 where for the $u^{th}$ asset $r_u$ is the expected return and $c_{u,v}$ is the historical price correlation between assets $u, v$. 
 \par 
 In Eq.~(\ref{eq:MPO_classic}), the first term represents maximization of the expected returns over the available assets. There are many methods for forecasting expected returns, e.g., based on market price, expert judgement, and historical price data \cite{huang2012mean, martin2017expected}. For simplicity, we model expected returns 
 as
 \begin{equation}
    r_{u} = p_w \bar a_u 
\end{equation}
where $\bar a_u$ is the average of $a_{u}$, the history of price data for the $u^{th}$ asset. The first constraint in Eq.~(\ref{eq:MPO_classic}) places a hard constraint on the total allocation of assets to sum to $b$. We emphasize that this constraint penalizes portfolios that do not allocate the entire budget as well as those that over commit. Finally, the second constraint accounts for diversification by asserting that the sum of covariance between asset prices $c_{u,v}$ be less than or equal to the risk threshold $c$. The historical price covariance is calculated as the correlation between pairs of assets by comparing the $p_w$ fraction of each asset's historical price data. Here covariance is defined as
\begin{equation}
\begin{aligned}
c_{u, v} = \frac{p_w^2 \sum^{N_f}_{l = 1}( a_{u,l}-  \bar a_u)( a_{v,l}- \bar a_v)}{N_f -1}
\end{aligned}
\end{equation}
where $a_{u,l}$ is the $l^{th}$ historical price value for asset $u$ and $N_f$ is the number of price points in the historical data. 
\par
We solve this variation of Markowitz portfolio selection using quantum annealing by casting the formulation in Eq.~(\ref{eq:MPO_classic}) into quadratic unconstrained binary optimization (QUBO).
We express the integer variable $z_u$ as a $w$-bit binary expansion 
 \begin{equation}
     z_u = \sum_{k = 1}^{w}{2^{k-1} x_{i(u,k)}}
     \label{eq:convert}
 \end{equation}
with $x_i \in \{0,1\}$ and the composite index $i(u,k) = (u-1)w + k$. The expected returns are then expressed as
\begin{equation}
    {r_{u} z_{u}} = { \sum_{k=1}^{w} 2^{k - 1} r_u x_{i(u,k)}}
\end{equation}
while the allocation constraint becomes the penalty term
\begin{align}
-\big(\sum_{u=1}^{m} \sum_{k=1}^{w} 2^{k - 1} p_w b x_{i(u,k)} - b \big)^2
\label{eq:norm_prices}
\end{align}
We consider a correlation threshold $c=0$ such that the correlation constraint becomes
\begin{align}
\sum_{u,v}^{m} c_{u, v} z_u z_v  =  \sum_{u,v}^{m} \sum_{k, k^{\prime}}^{w}  2^{k - 1} 2^{k^{\prime} - 1} c_{u,v} x_{i(u,k)} x_{ j(v,k^{\prime})}.
\label{eq:norm_prices}
\end{align}
Our formulation of Markowitz portfolio selection as an unconstrained optimization problem then becomes

\begin{maxi}
  {x}{{\theta_1 \sum^{n}_{i} r_{i}  x_{i}} 
\breakObjective{- \theta_2(\sum^{n}_{i}2^{k - 1}b p_w   x_{i} - b)^2}
\breakObjective{- \theta_3 \sum^{n}_{i, j} c_{i, j} x_{i} x_{ j}}}{}{}
{\label{eq:MPO_unconstrained}}
\end{maxi}
where the problem size $n = mw$, $r_i = 2^{k - 1} r_u$, $c_{i, j} = 2^{k - 1} 2^{k^{\prime} - 1}  c_{u,v}$, and  $\theta_1, \theta_2$ and $\theta_3$ are Lagrange multipliers used to weight each term for maximization or penalization. 
\par
For purposes of benchmarking, we generate an ensemble of problem instances by sampling from uniform random price data with a seed of ${b}/{5}$ . A random number is drawn as the initial price $a_{u,1}$ and every subsequent historical price point up to the purchasing price is $-25\%$ to $+25\%$ of the previous price $a_{u, l}$. The price range was set to be between $b/10$ and $b$ with $N_f = 100$ historical price points per asset. In addition, we normalize all $a_{u, l}$ by $a_{u, N_f}$ to keep all asset prices to a similar range. 
\par 
We set $\theta_1 = 0.3, \theta_2 = 0.5, \theta_3 = 0.2$ in the problem instances  where $\theta_2$ is set higher to enforce the budget constraint. These weights were chosen after testing which combination stayed on budget and gave some diversity. By keeping $\theta_2$ constant and increasing $\theta_3$ while decreasing $\theta_1$, an investor could increase the diversity relative to the potential returns and vice versa when decreasing $\theta_3$ relative to $\theta_1$. We generate $1000$ problems for each problem size with $m = 2, 3, 4, 5$ assets and $w = 4$ slices.
\subsection{QUBO to Ising Hamiltonian \label{sec:qubo_ising}}
We formalize the unconstrained portfolio optimization problem in Eq.~(\ref{eq:MPO_unconstrained}) to quadratic unconstrained binary optimization (QUBO) as 
\begin{mini}
{x}{\Big(\sum_{i}^{n} q_{i} x_{i} + \sum_{i, j}^{n} Q_{i, j} x_{i} x_{ j} + \gamma\Big)}{}{}
{\label{eq:QUBO}}
\end{mini}
where $q_i$ is the linear weight for the $i^{th}$ spin, $Q_{i,j}$  is the quadratic weight for interactions between the $i^{th}$ and $j^{th}$ bits, and  $\gamma$ is a constant. Note that our definition of QUBO expresses optimization as minimization by switching the sign of Eq.~(\ref{eq:MPO_unconstrained}) to be consistent with the use of quantum annealing to recover the lowest-energy state. The corresponding relationships with the original problem instance are given as
\begin{equation}
\begin{aligned}
& q_i = -\theta_1 r_{i} - 2 \theta_2 b^2 p_w  \\
& Q_{i,j} = \theta_2 b^2 p_{w}^2  +  \theta_3 c_{i,j}  \\
& \gamma = \theta_2 b^2
\end{aligned}
\end{equation}
Similarly, the quadratic binary form may be reduced to a classical Ising Hamiltonian 
\begin{equation}
\label{Ising Hamiltonian}
H(s) =  \sum_{i}  s_i h_i + \sum_{i,j}s_i s_j J_{ij} + \beta
\end{equation}
where spin $s_i\in \{-1, 1\}$ is defined by $s_i = 2 x_1 - 1$ with $s = (s_1, s_2, \ldots, s_n)$ while $h_i$ is the spin weight, $J_{ij}$ is the coupling strength, and $\beta$ is a problem-specific constant. The parameters for the Ising Hamiltonian are given as 
\begin{equation}
\begin{aligned}
& J_{i,j} = \frac{1}{4} Q_{i,j}\\
& h_i = \frac{q_i}{2} + \sum_{j} J_{i,j}\\
& \beta = \frac{1}{4} \sum_{i,j} Q_{i,j}  + \frac{1}{2} \sum_{i} q_i + \gamma
\end{aligned}
\end{equation}
The classical Ising formulation is then converted into a corresponding quantum Ising Hamiltonian given by Eq.~(\ref{eq:Ising_Hamiltonian}) using the correspondence $s_i \rightarrow \sigma_{i}^{z}$.

\subsection{Computational Methods}
We used a D-Wave 2000Q quantum annealer for our experiments. We calculate the probability of success, the probability of chain breaks, and the energy distribution across each problem instance. For each instance, we estimated these metrics by collecting $N_s=1000$ samples of the computed solution. We used D-Wave's solver API (SAPI) with Python $2.7$ to solve each instance of Markowitz portfolio selection using the hardware controls outlined in Sec.~\ref{sec:Controls}. We ran $1,000$ samples per problem over a set of $1,000$ problems for forward annealing examples an $100$ problems for revere annealing examples. We implement the majority vote post-processing technique for any broken chains to interpret raw solutions returned by the $2000Q$. The program implementation and data sets collected from these experiments are available online \cite{repository}.
\par
For benchmarking purposes, we also solved each problem instance using brute force search for the minimal energy solutions of the QUBO formulation. We computed the complete energy spectrum for each portfolio instance. These energy spectrum and the corresponding states were then used as ground truth for testing the accuracy of results obtained from quantum annealing. By sorting the spectrum, we benchmarked the success of reverse annealing using initial states $e_{i}$ sampled from these different parts of the spectrum.
\section{Results \label{sec:Results}}
\begin{figure}[h!]
\centering
\includegraphics[width=85mm]{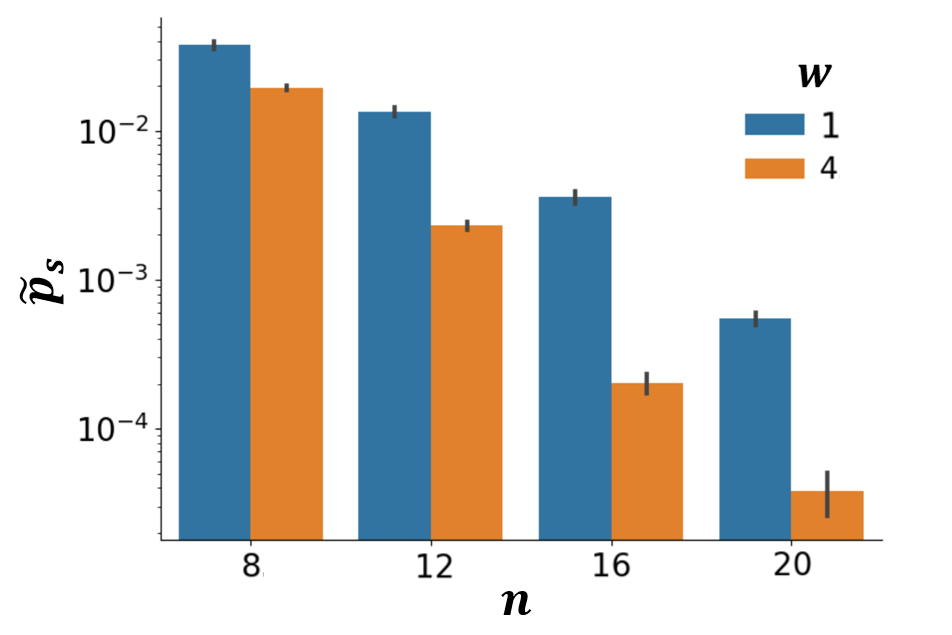}
\caption{The average probability of success over $1000$ problems each with $1000$ samples using CMR, $g= 0$, and $ T= 15$ $\mu$ s. The comparison is between a set of problems from problem sizes $8$ to $20$ for $w = 1$ (yellow) and $w = 4$ (blue). The problems set to slices $w = 1$ are much less complex and therefore have a much higher probability of success.}
\label{fig:pos_slices}
\end{figure}

We benchmark quantum annealing controls by evaluating their influence on the probability of success and probability of chain breaks across problem instances. We first characterize how problem parameters influence the baseline performance by estimating the probability of success for forward annealing using $T = 15~\mu s$, $g = 0$, and a randomized embedding strategy. As shown in Fig.~\ref{fig:pos_slices}, we compare $\tilde{p}_{s}$ for two cases of $w = 1$ and $w = 4$ across increasing $n$. The estimated probability of success for problems with $w = 4$ is consistently higher for problems with no slicing.
\begin{figure}[h!]
\centering
\includegraphics[width=85mm]{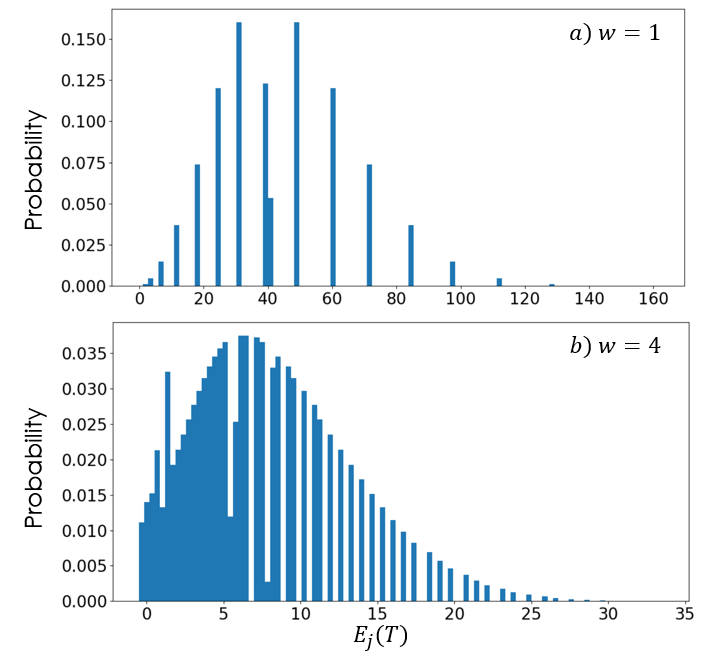}
\caption{Probability histogram ($100$ bins) of all possible energies for problem of size $20$ where $a)$ is of $w = 1$ and $b)$ is of $w = 4$. There is a higher density of states close to $e_{0}$ in figure $b)$ and therefore more opportunities to jump to an excited state throughout the sample.} 
\label{fig:slices_hist}
\end{figure}
\par
These results are explained by the energy spectra for the different problem parameters, which indicate sharp differences in the density of states.  As shown in Fig.~\ref{fig:slices_hist}, a typical problem instance with $w = 4$ has a much higher density of states than those with no slicing ($w=1$). Intuitively, the single-slice behavior results from the specification that the price for each asset is proportional to budget, and, therefore, only a single asset may be selected without penalty when $w = 1$. However, the number of satisfying solutions $v$ increases for arbitrary $w$ combinatorially and, as shown in  Appendix~\ref{appx:num_solutions},
\begin{equation}
\begin{aligned}
v = \frac{(2^{w-1} + m - 1)!}{(2^{w-1})! (w - 2)!}.
\end{aligned}
\label{num_solutions}
\end{equation}
Consequently, the probability to recover the lowest-energy state competes with these closely spaced, higher energy solutions, which leads to a corresponding decrease in the probability of success. For the remaining benchmark tests below, we chose $w = 4$ as it represents a more challenging test for the quantum annealer as well as a greater interest to real-world financial applications. 
\subsection{Benchmarking Forward Annealing Controls}
\subsubsection{Embedding \label{sec:Embedding}}
Embedding generates and places the physical spin chains for each logical spin on the quantum annealing hardware. We evaluated the CMR and clique embedding algorithms described in Sec.~\ref{sec:Embedding} by estimating the probability of success across problem sizes of $m = 8, 12, 16,$ and $20$ logical spins. For all problem instances of a same problem size, we use the same embedding because they require the same number of fully connected logical spins. We set the parameters of the embedded Ising Hamiltonian by scaling the inter-chain couplings $J_{i,j}$ to lie in the range $[-1,+1]$. We scale all $J_{i,j}$ using a rescale factor of $\frac{1}{j_{max}}$ where $j_{max}$ is the largest $J_{i, j}$ so all embedded $J_{i, j}= \frac{1}{j_{max}} J_{i, j}$. This scales all $J_{i, j }$ to be between $+- 1$. The intra-chain coupling strength is set to $-1$ to have a negative bias stronger than the $J_{i,j}$ values which range $-10^{-1} \leq J_{i,j} \leq 10^{-1}$ due to our data generation and normalization techniques. 
\par
\begin{figure}[h!]
\centering
\includegraphics[width=85mm]{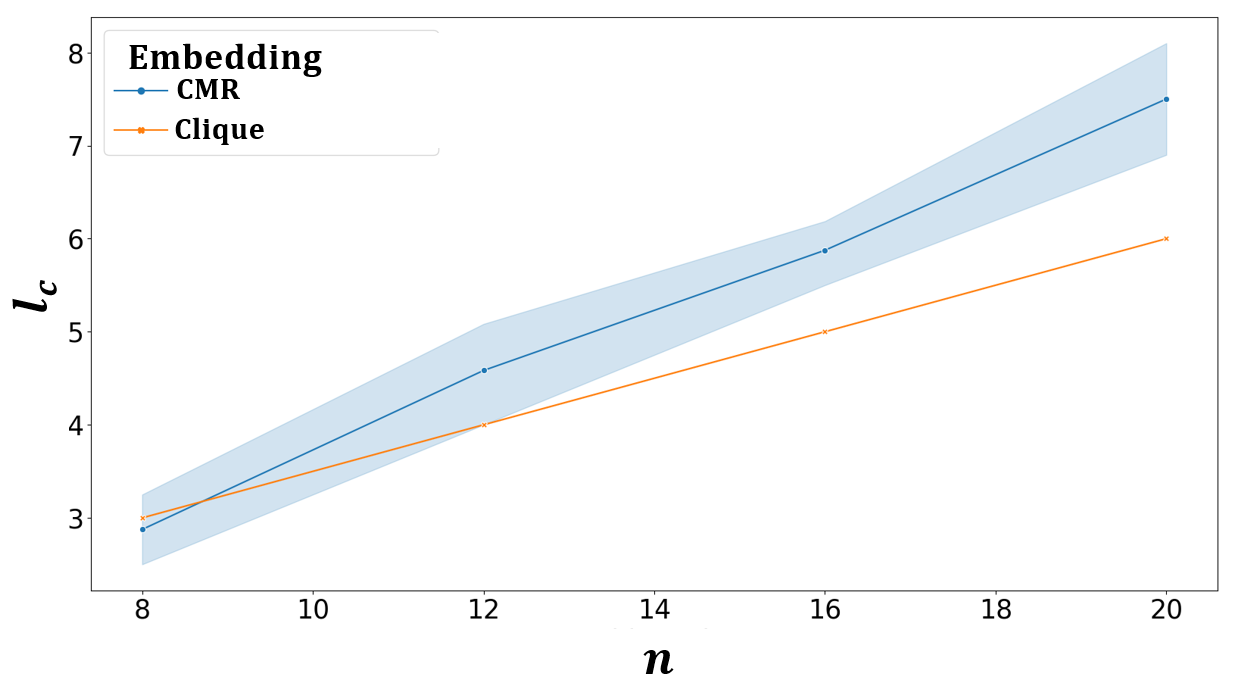}
\caption{The average chain length over all chains for a given embedding clique and CMR embedding as $n$ increases.} 
\label{fig:embedding_size}
\end{figure}

The average chain length $\langle l_c \rangle$ from CMR and clique embedding methods grows with the number of logical spins $n$. The average is computed with respect to all chains in an embedding and plotted with respect to $n$ in Fig.~\ref{fig:embedding_size}. As expected by Eq.~(\ref{eq:clique_chain_length}), the clique embedding method has a uniform chain length for each $n$. By contrast, the CMR method generates chains of variable length as indicated by the the average chain length and variance shown in the plot. 
\par
\begin{figure}[h!]
\centering
\includegraphics[width=85mm]{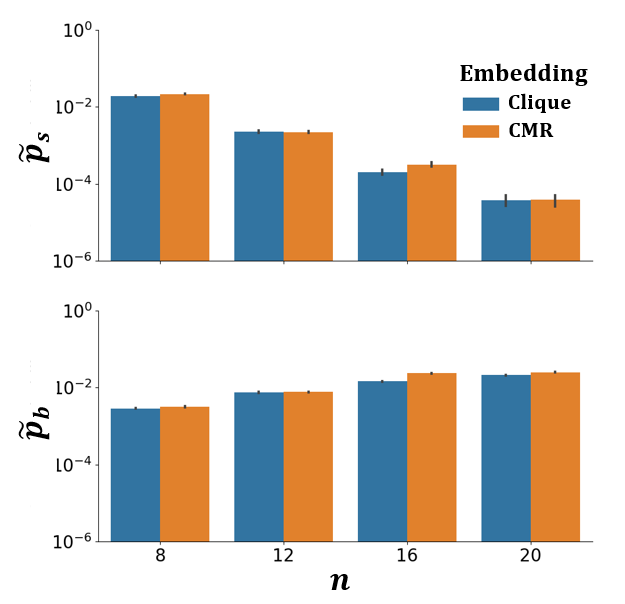}
\caption{The $\tilde{p}_{s}$ (top) and $\tilde{p}_{b}$(bottom) on a log scale over  $1,000$ samples for $1,000$ problems comparing CMR to clique embedding for parameter settings of $g = 0$ and $T = 100~\mu s$.}
\label{fig:pos_embedding}
\end{figure}

From each of the embedding methods, we estimate the probability of success and probability of broken chains. As shown in Fig.~\ref{fig:pos_embedding}, we observe very small differences in both metrics with increasing problem size. From fitting the resulting point to an exponential, we find $\tilde{p}_{s}$ decays sub-exponentially with respect to $n$ with rate $-0.523$ for the CMR embedding and rate $-0.528$ for the clique embedding. We find that $\tilde{p}_{b}$ grows at a sub-exponential rate of $0.1824$ for CMR embedding and $0.1656$ for clique embedding as $n$ increases. There is not a significant difference in the $\tilde{p}_{s}$ performance between CMR and clique embedding, but clique embedding requires a fewer number of spins as $n$ increases and shows a slight improvement in $\tilde{p}_{b}$. Therefore, we chose to use clique embedding for subsequent benchmarks.
\subsubsection{Forward Annealing Time}
\begin{figure}[h!]
\centering
\includegraphics[width=\columnwidth]{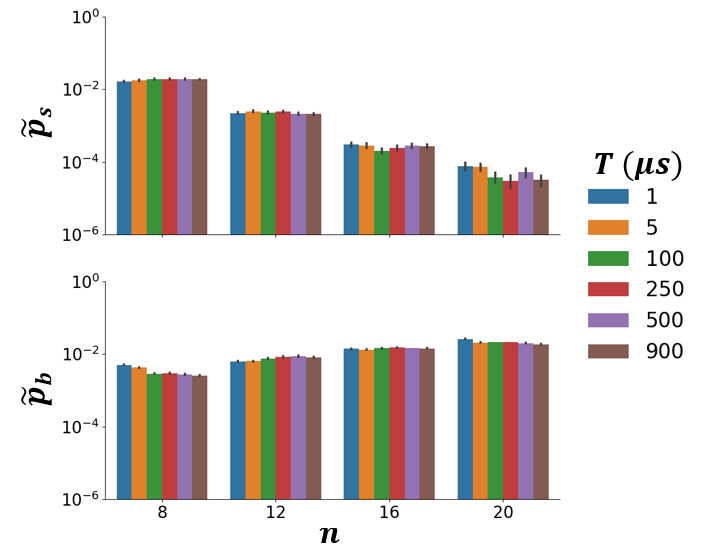}
\caption{The average $\tilde{p}_{s}$ (top) and $\tilde{p}_{b}$(bottom) on a log scale over $1000$ samples for $1000$ problems at various annealing times for parameter settings of $g = 0$ and clique embedding.}
\label{fig:anneal_times}
\end{figure}

According to the adiabatic theorem, forward annealing more slowly should increase the probability of the system remaining in the ground state and thus increase the probability of success. We varied the forward annealing time $T$ from $1~\mu s$ to $999 \mu s$, which is the broadest range accessible on the D-Wave $2000Q$. As shown in the upper panel of Fig.~\ref{fig:anneal_times}, we observed statistically insignificant changes in the probability of success as annealing time increased at each problem size. Fitting the average probability of success with respect to problem size for the annealing time $T = 100~\mu s$, yields a sub-exponential decay rate for $\tilde{p}_{s}$ given by $-0.528$ and a sub-exponential growth rate for $\tilde{p}_{b}$ given by $0.1628$ as $n$ increases. We do observe a statistically significant difference in the estimated probability of chain breaks $\tilde{p}_{b}$ with respect to forward annealing time as shown in the lower panel of Fig.~\ref{fig:anneal_times}. For $T=100~\mu s$, we recover a growth rate of $0.1656$ for the probability of chain breaks with respect to problem size.
\subsubsection{Spin Reversal \label{sec:Spin_Reversal}}
\begin{figure}[h!]
\centering
\includegraphics[width=85mm]{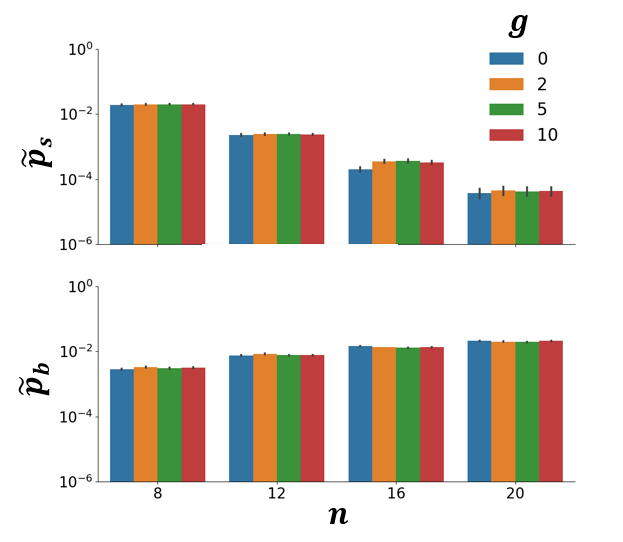}
\caption{The $\tilde{p}_{s}$ (top) and $\tilde{p}_{b}$ (bottom) on a log scale over $ N_s = 1000$ samples for $N_p = 1000$ problems at $g = 0 \rightarrow 10$) for parameter setting of $T = 100 \mu s$ and clique embedding.} 
\label{fig:sr}
\end{figure}

As discussed in Sec.~\ref{sec:Embedding}, embedding maps a logical spin to many physical spins by creating strongly coupled chains. Coupling of these embedded spins via $J_{i,j}$ in Eq.~(\ref{eq:Ising_Hamiltonian}) can lead small biases that may be amplified by imperfections in the hardware. A spin reversal transform mitigates against bias errors by reversing the sign of a spin in the Ising Hamiltonian. This transform preserves the logical problem but reverses the bias error on the embedded spin chain. By randomly selecting a subset of spins to revise, we evaluate the influence of spin-reversal transform on the probability of success. We use $g$ transforms when estimating the probability of success for a given problem instance, such that there are ${N_s}/{g}$ samples per transform. We observed nominal improvements in Fig.~\ref{fig:sr} by using at least $g = 2$ with no advantage to using $g > 2$. For $g = 2$, we observe an sub-exponential decay rate of $-0.505$ for $\tilde{p}_{s}$ and a sub-exponential growth rate of $0.146$ for $\tilde{p}_{b}$ as problem size increases. 

\subsection{Benchmarking Reverse Annealing Controls}
From the reverse annealing controls listed in Sec.~\ref{sec:Control_AnnealTime}, we designed three experiments based on the $e_{i}$ for the reverse annealing heuristic that include (\textit{i}) starting in the known ground state $e_{0}$,  (\textit{ii}) starting in the known first excited state $e_{1}$, and  (\textit{iii}) starting in the lowest-energy state obtained from $1000$ forward annealing samples $e_{f}$. We then sweep over various schedules to find the optimal $s_p$ with a range of $[0.1, 0.9]$ and $t_p$ with a range of $[15 \rightarrow 800]\mu s$. The $t_r$ and $t_q$parameters were set to be constant and symmetric at $5 \mu s$ each. Thus, the total anneal time is $T^{\prime} = t_r + t_p + t_q$ where $t_p$ is the time parameter that we chose to analyze. For all experiments, we ran the reverse annealing iterative heuristic with $1000$ samples for $100$ random problems were also used in the forward annealing experiments. We estimated the probability of success for reverse annealing with respect to different choices for $e_{i}$, $s_p$, and $t_p$. We compared the combined heuristic of reverse annealing with forward annealing to forward annealing alone with $\tilde{p}_{s}$, $\tilde{p}_{b}$,  as well as the frequency of finding energies in excited states to forward annealing alone \footnote[1]{After completing the majority of experiments on the D-Wave processor DW\_2000Q\_2\_1, the remaining experiments were performed on D-Wave processor DW\_2000Q\_5. This included the parametric tests of reverse annealing with respect to $s$ and $t_p$. Prior to testing, we confirmed computational consistency between the results generated using the first device and those using the second. We evaluated differences in $\tilde{p}_s$ and standard deviation between the processors by comparing a previous reverse annealing experiment on the DW\_2000Q\_2\_1 to the same experiment on the DW\_2000Q\_5. We found that the same $\tilde{p}_s$ using both devices and a standard deviation that was within $10^{-5}$ of the measurements on the previous D-Wave processor.}. 
\par
By setting $e_{i}$ to the ground state, we tested for parameters $s_p$ and $t_p$ that decrease $\tilde{p}_s$ when the quantum annealer is fed the correct solution. For this experiment, $\tilde{p}_s$ can be thought of as the probability of staying in $e_{0}$
\begin{equation}
 \tilde{p}_s(e_{0} \rightarrow e_{0}) = p_f * \tilde{p}_{s}
\end{equation}

\begin{equation}
p_f * \tilde{p}_s = \frac{\sum_i^{N_p} \alpha_i}{N_p} *  \frac{\sum_i^{N_p}\sum_{j}^{N_s}\delta_{ij}}{N_s}
\end{equation}
\par
where $p_f$ is the probability that forward annealing found the ground state, $\alpha_i \in \{0, 1\}$ indicates whether forward annealing found the ground state for the $i^{th}$ problem prior to reverse annealing, and $\delta_j \in \{0, 1\}$ is a variable indicating whether the $j^{th}$ sample of the $i^{th}$ problem was measured to be the ground state with reverse annealing. By setting $e_{i} = e_{1}$, we tested whether reverse annealing enhances the probability to populate the ground state. For these tests, $\tilde{p}_s$ estimates the probability of moving from an excited state to the ground state
\begin{equation}
\tilde{p}_s(e_{e} \rightarrow e_{0}) = (1 - p_f) * \tilde{p}_{s}
\end{equation}
\begin{equation}
(1 - p_f) * \tilde{p}_{s} = \frac{\sum_i^{N_p} (1 - \alpha_i)}{N_p} *  \frac{\sum_i^{N_p}\sum_{j}^{N_s}\delta_{ij}}{N_s}.
\end{equation}
In addition to testing reverse annealing at $e_{i} = e_{0}$ and $e_{1}$, We tested reverse annealing in combination with forward annealing for which $\tilde{p}_s$ estimates the cumulative probability of finding the correct solution state.
\begin{equation}
\tilde{p}_{s}(R) = \tilde{p}(e_{0} \rightarrow e_{0}) + \tilde{p}(e_{e} \rightarrow e_{0})
\end{equation}
For these experiments, we found it useful to primarily analyze $\tilde{p}_{s}(R) - \tilde{p}(e_{0} \rightarrow e_{0}) = \tilde{p}(e_{e} \rightarrow e_{0})$ to determine if reverse annealing improved upon the $\tilde{p}_{s}$ of forward annealing. 
\par
\begin{figure}[h!]
\centering
\includegraphics[width=85mm]{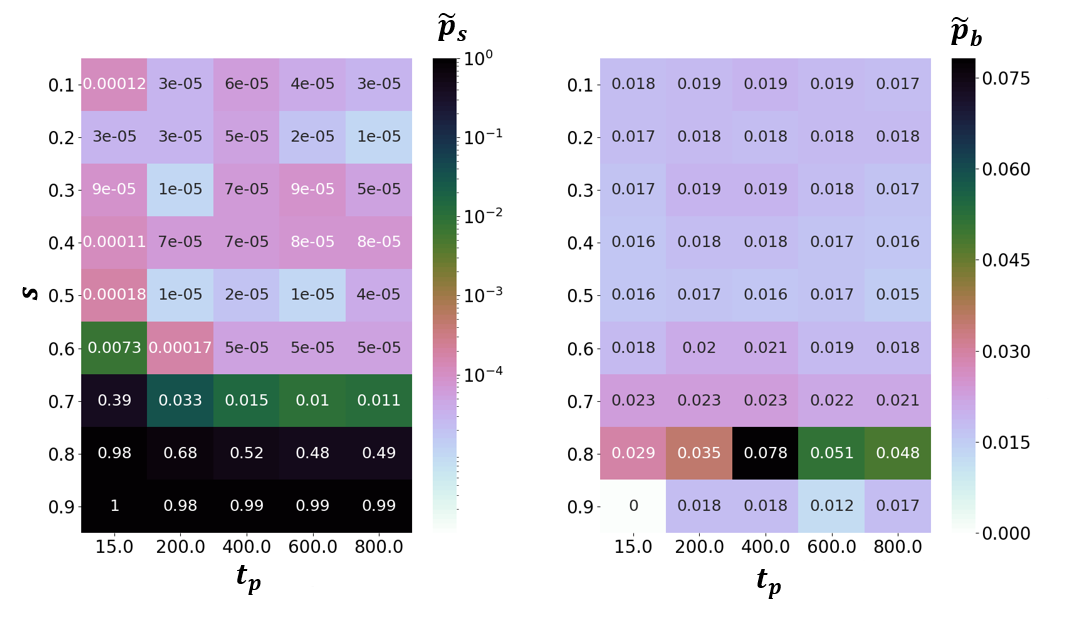}
\caption{The $\tilde{p}_{s}$ (left) and $\tilde{p}_{b}$ (right) for reverse annealing where $e_{i} = e_{0}$ and as $s = [0.1 \rightarrow 0.9]$ and $t_p = [15\mu s \rightarrow 800 \mu s]$ for $n = 20$ with $m = 5$ assets and $w = 4$. } 
\label{fig:pos_ground}
\end{figure}

The results from setting $e_{i} = e_{0}$ for each problem  with a problem size of $n = 20$ where $m = 5$ and $w = 4$ is shown in Fig.~\ref{fig:pos_ground}. Because the computation begins in the correct solution state, this test measures the probability by which reverse annealing introduces errors into the correct solution. Ideally, $\tilde{p}_s$ will remain near unity for all $s_p$ and $t_p$. We observe that reverse annealing causes the system to leave the ground state with $\tilde{p}_{s}$ reducing to on the order of $10^{-5}$ by annealing back to at least $s = .6$ and increasing $t_p \geq 200 \mu s$ 
\par
\begin{figure}[h!]
\centering
\includegraphics[width=85mm]{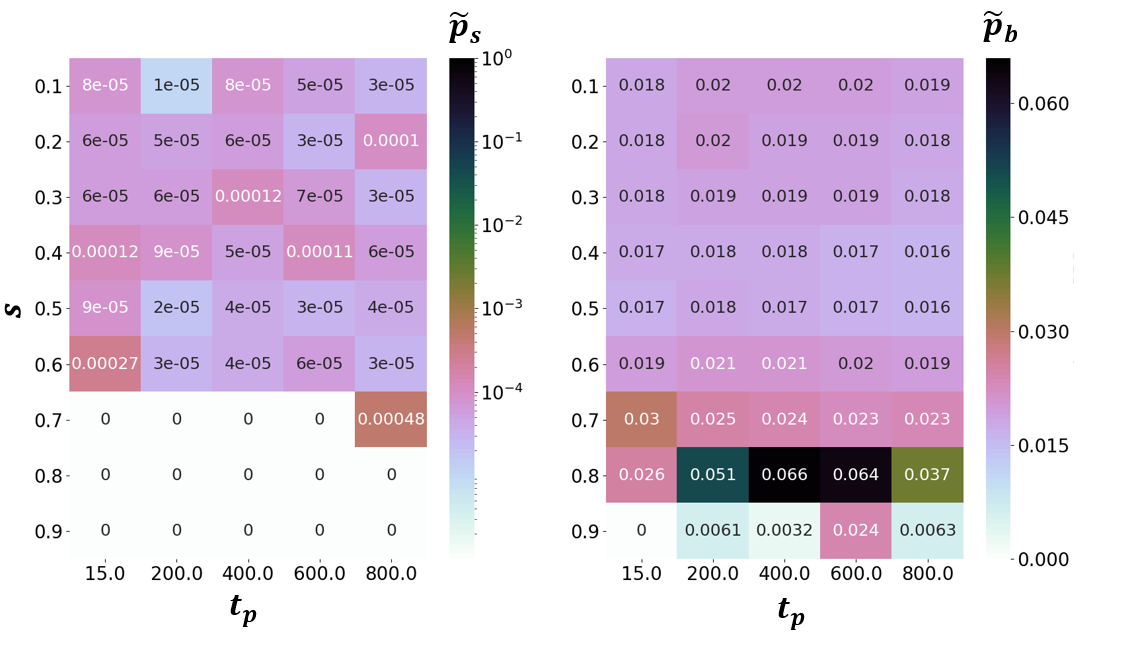}
\caption{The $\tilde{p}_{s}$ (left) and $\tilde{p}_{b}$ (right) for reverse annealing where $e_{i} = e_{1}$ for each problem, $s = [0.1 \rightarrow 0.9]$, and $t_p = [15 \mu s \rightarrow 800 \mu s]$ for problem size $20$ with $5$ assets and $4$ slices.} 
\label{fig:pos_excited}
\end{figure}

The results from setting $e_{i} = e_{1}$ with a problem size of $n = 20$ where $m = 5$ and $w = 4$ for each problem is shown in Fig.~\ref{fig:pos_excited}. A maximal value of $4.8\times 10^{-4}$ for $\tilde{p}_s$ is found with parameters $s = 0.7$ and $t_p = 800~\mu s$. This is is a $\tilde{p}_s$ one order of magnitude higher than what is observed with forward annealing. This suggests that if $e_{i}$ is very close to $e_{0}$, there may be some benefit to choosing reverse annealing over forward annealing. 
\par
\begin{figure}[h!]
\centering
\includegraphics[width=85mm]{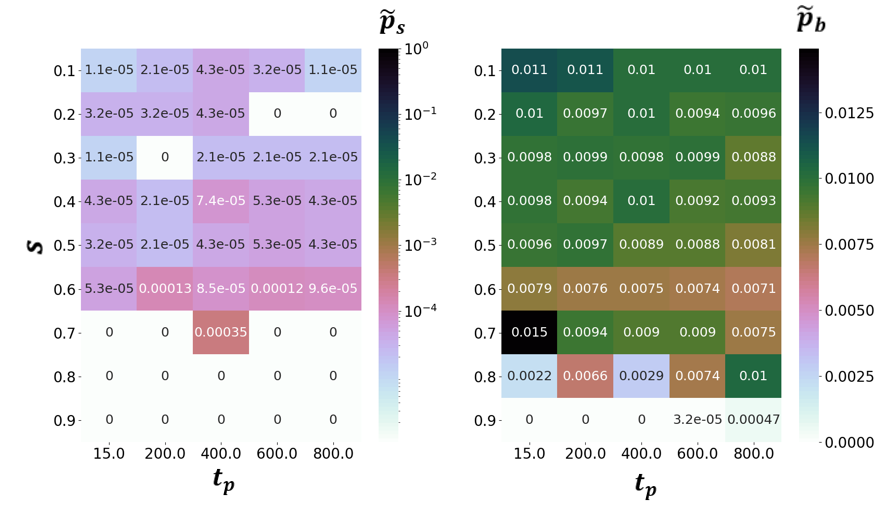}
\caption{The $\tilde{p}_{s}$ (left) and $\tilde{p}_{b}$ (right) for reverse annealing where $e_{i} = e_{f}$ for each problem, $s = [0.1 \rightarrow 0.9]$, and $t_p = [15 \mu s \rightarrow 800 \mu s]$ for problem size $20$ with $5$ assets and $4$ slices. The $6$ problems where $e_{f} = e_{g}$ were excluded. Thus, $\tilde{p}_s = p(e_{e} \rightarrow e_{0})$.} \label{fig:pos_no_ground}
\end{figure}

When solving optimization problems for applications in practice, the ground state and excited state will be unknown. However, one approach is to use reverse annealing in addition to forward annealing by using the lowest energy state found with $1,000$ forward annealing samples $e_{f}$ as $e_{i}$ for another $1,000$ samples of reverse annealing. The next experiment tests whether reverse annealing used in combination with forward annealing increases $\tilde{p}_s$ with a problem size of $n = 20$ where $m = 5$ and $w = 4$ . The experimental results from setting $e_{i} = e_{f}$ is shown in  Fig.~\ref{fig:pos_no_ground}. These tests were constructed to determine when combining reverse annealing with forward annealing can improve upon forward annealing. Therefore, we removed the 6 problems forward annealing provided an $e_{i} = e_{0}$ and thus $\tilde{p}_{s}$ for this experiment is given by $\tilde{p}_s(R) - p(e_{0} \rightarrow e_{0})$ in this analysis. Similar to the previous experiment in Fig.~\ref{fig:pos_excited}, the $\tilde{p}_{s}$ is at best on the order of $10^{-4}$ at parameters $s = 0.7$ and $t_p = 400 \mu s$ which is one order of magnitude greater than the forward annealing experiments.
\par
\begin{figure}[h!]
\centering
\includegraphics[width=85mm]{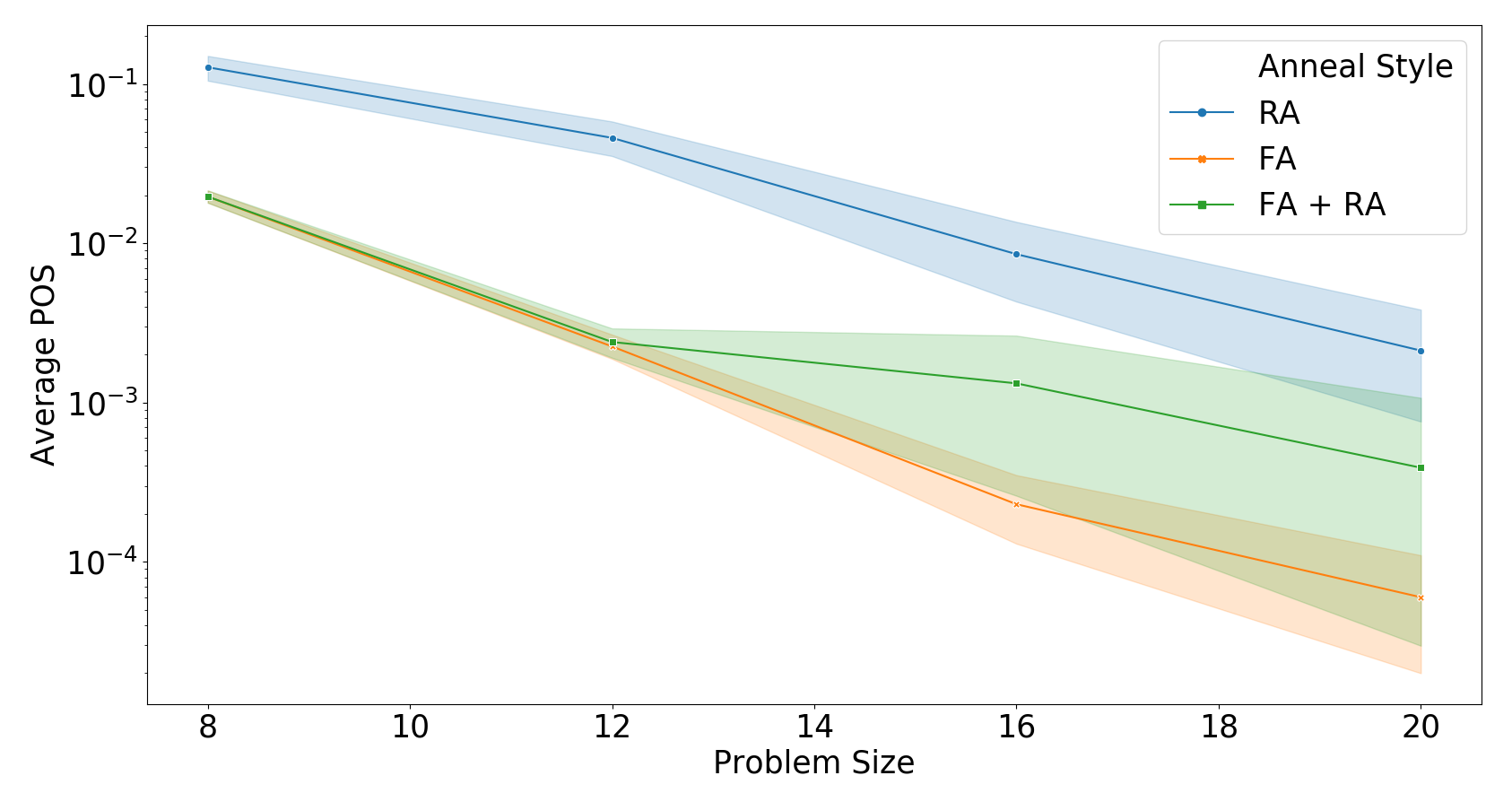}
\caption{The $\tilde{p}_s$ as a function of $n$ over a set of $100$ problems each with $1000$ samples. We compare reverse annealing (blue) with $e_{i} = e_{f}$, $s = .7$, and $t_p = 400 \mu s$ to forward annealing (orange) with clique embedding, $g = 0$, and annealing time $= 100$ $\mu s$. We also compare the combination of forward annealing and reverse annealing where the $\tilde{p}_{s}$ is chosen by problem (green). In this green trend, the $\tilde{p}_s$ is calculated using the forward annealing $\tilde{p}_{s}^{(k)}$ for the $6$ problems where forward annealing would have provided reverse annealing with an $e_{i} = e_{0}$ and the reverse annealing $\tilde{p}_{s}^{(k)}$ for the $94$ problems where $e_{i} \neq e_{0}$. } 
\label{fig:pos_track}
\end{figure}

\begin{figure}[h!]
\centering
\includegraphics[width=85mm]{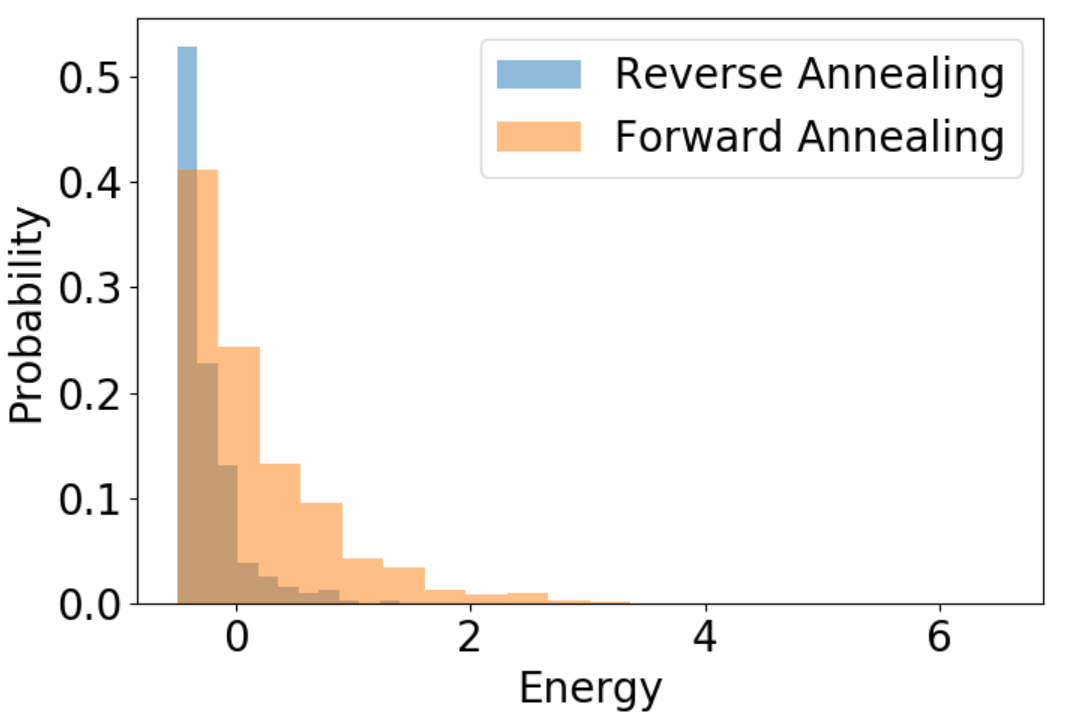}
\caption{A probability histogram ($20$ bins) comparing all energies found with forward annealing and reverse annealing from all $1000$ samples for the $94$ problems where $e_i \neq e_0$ for problems with $m = 5$ assets and $w = 4$ .} 
\label{fig:ra_vs_fa}
\end{figure}

 
Fig. ~\ref{fig:pos_no_ground} shows a potential for reverse annealing to improve upon results found with forward annealing in $\tilde{p}_{s}$. Therefore, we take a set of $100$ problems solved with reverse annealing and forward annealing and compare the $\tilde{p}_{s}$ of forward annealing (orange) alone to the $\tilde{p}_{s}$ of reverse annealing alone (blue) to the $\tilde{p}_{s}$ with a selection of either forward annealing or reverse annealing (green). If for a problem forward annealing found at least one ground state the forward annealing $\tilde{p}_{s}$ was plotted for that problem ($6$ problems) and otherwise the reverse annealing $\tilde{p}_{s}$ was plotted ($94$ problems). The  $\tilde{p}_s$ is measured over $n$ ranging from $[8, 20]$. The reverse annealing parameters are set to have an $e_{i} = e_{f}$ , $s = .7$, and $t_p = 400 \mu s$. As shown in Fig.~\ref{fig:pos_track}, we observe that when taking the combination of best results from forward annealing and reverse annealing with $e_{i} = e_{f}$, we get a $\tilde{p}_{s}$ that improves by an order of magnitude over forward annealing alone for $n = [16, 20]$ with a sub-exponential decay at a rate of $-0.309$. Note that although the blue reverse annealing trend looks to perform the best, however this trend is artificially inflated because $6$ of the problems have $e_{i} = e_{0}$ which has been demonstrated in Fig.~\ref{fig:pos_ground} to yield a $\tilde{p}_s$ on the order of $10^{-2}$ at $s = .7$ and $t_p = 400 \mu s$. 
We next visualize a histogram, as seen in Fig.~\ref{fig:ra_vs_fa}, of all energies recorded from $1000$ samples returned for a set of $94$ problems where forward annealing did not find $e_{0}$ with $n = 20$. We compare forward annealing to reverse annealing where $e_{i} = e_{f}$. We observe even for problems where neither reverse annealing or forward annealing found $e_{0}$, reverse annealing still on average finds a lower energy solution than forward annealing. 
\section{\label{sec:level1} Conclusions}
We have benchmarked quantum annealing using Markowitz portfolio selection to evaluate the effects of various controls on probability of success and chain breaks. 
We have explored a variety of quantum annealing controls including the embedding algorithm, the forward annealing time $T$, and the number spin reversal transforms $g$. When comparing clique embedding against CMR embedding, we found little difference in the estimated probability of success $\tilde{p}_{s}$ as both techniques yielded a sub-exponential decay for $\tilde{p}_s$ with exponents of $-0.528$ and $-0.523$, respectively. We did observe that CMR demonstrated a slightly higher probability of chain breaks $\tilde{p}_{b}$, and we considered this a sufficient justification to use the clique embedding for studying the fully connected problems Markowitz portfolio selection problem. 
\par 
When varying the forward annealing time $T \in [1 \mu s, 999 \mu s]$, we found that $\tilde{p}_b$ was slightly higher in the range $T = [1 \mu s, 5 \mu s ] $ while increasing the annealing time further yielded little to no improvement. For this reason, we chose to continue all future forward annealing experiments using $T = 100 \mu s$  where the exponential decay rate in $p_s$ was $-0.528$. When varying $g = [0, 10]$, we found small improvements in $\tilde{p}_s$ between $g = 0$ and $g = 2$ where the exponential decay rate became $-0.505$ without much change from increasing the value of $g$ further, and there was no consistent difference in $\tilde{p}_b$.
\par
We benchmarked reverse annealing controls with respect to the parameters $e_{i}$, $s$, and $t_{p}$. We began by observing the results in $\tilde{p}_{s}$ and $\tilde{p}_b$ at  $n=20$. We consistently observed that $\tilde{p}_b$ was the same order of magnitude as with the forward annealing experiments and $\tilde{p}_b$  was consistently highest for $s = 0.8$. By setting $e_{i} = e_{0}$, we observed that the $\tilde{p}_s$ decreases exponential as $s$ increased.  By setting $e_{i} = e_{1}$, we observed that reverse annealing had a $\tilde{p}_s$ an order of magnitude higher than forward annealing. From these results, we conclude that when $e_{i}$ is close to the ground state, reverse annealing provided some advantage over forward annealing. Because in general the ground state won't be known for a problem, we developed a heuristic which sets $e_{i} = e_{f}$ where we again observed $\tilde{p}_s$ to be an order of magnitude higher than using forward annealing alone. 
\par 
We further evaluated $\tilde{p}_s$ as a function of $n$ to compare reverse annealing with  $e_{i} = e_{f}$, $s = 0.7$, and $t_p = 400 \mu s $ to forward annealing with clique embedding, $T = 100 \mu s $, and $g =0$ alone. In particular, we used the $\tilde{p}_{s}^{(k)}$ of forward annealing for the $6$ problem instances in which $e_{i} = e_{0}$ and the $\tilde{p}_{s}^{(k)}$ of reverse annealing for the $94$ problems where $e_{i} \neq e_{0}$.  We continued to observe reverse annealing demonstrate an order of magnitude increase in $p_s$ over forward annealing alone. Lastly, by creating a histogram which plots the lowest energies found across $1000$ samples for the $94$ problems where $e_{i} \neq e_{0}$, we found that reverse annealing$(e_{i} = e_{f})$ on average finds lower energy solutions as compared to forward annealing. 
\par
In summary, the benchmarks presented here evaluate a variety of quantum annealing controls with respect to the baseline ground truth for portfolio selection. By comparing the observed influence of these controls on the performance of solution accuracy, we have developed insights into the best selections of controls for solving these problems with the highest accuracy which may help guide the future use of quantum annealing as a meta-heuristics for optimization.
\section*{\label{sec:ack} Acknowledgements}
This work is supported by the Department of Energy, Office of Science, Early Career Research Program. This research used quantum computing resources of the Oak Ridge Leadership Computing Facility, which is a DOE Office of Science User Facility supported under Contract DE-AC05-00OR22725. This manuscript has been authored by UT-Battelle, LLC under Contract No. DE-AC05-00OR22725 with the U.S. Department of Energy. The United States Government retains and the publisher, by accepting the article for publication, acknowledges that the United States Government retains a non-exclusive, paid-up, irrevocable, world-wide license to publish or reproduce the published form of this manuscript, or allow others to do so, for United States Government purposes. The Department of Energy will provide public access to these results of federally sponsored research in accordance with the DOE Public Access Plan. (http://energy.gov/downloads/doe-public-access-plan).
\bibliographystyle{unsrt}
\bibliography{main.bib}

\appendix




\section{Number of Combinations Constrained to the Budget \label{appx:num_solutions}}
Assuming the optimal solution lies where the total value of assets bought equals the budget, the number of solutions which need to be checked is drastically reduced. If we have 1 asset, the only solution is buying the slice equal to 1. If we have 2 assets, the slice of the 2nd asset is dictated by whichever slice is chosen from the 1st asset. If the number of slices chosen is $w$, then we know that the slices correspond to $1,\frac{1}{2},\frac{1}{4},\frac{1}{8}..\frac{1}{2^{w}}$. This gives a total of $2^{w}+1$ (since we can also buy 0 for all slices) which are less than or equal to the budget. Mathematically, this can be expressed as:

\begin{equation}
\begin{aligned}
\text{\# solutions} = \sum_{a_1=0}^{2^{w}}\sum_{a_2=0}^{2^{w}-a_1}1=2^{w}+1
\end{aligned}
\end{equation}

This is an equivalent problem to stating how many distinct terms are in the binomial $(a_1+a_2)^{2^{w}}$. 

Extending this to an arbitrary amount of assets $(m)$, this equates to finding how many distinct terms are in the multinomial expansion $(a_1+a_2+...+a_m)^{2^{w}}$ which can be found using the following equation

\begin{equation}
\begin{aligned}
\text{\# solutions} = \prod_{a=1}^{m-1}\frac{2^w+a}{a}=\frac{(2^w + m - 1)!}{(2^w)! (m - 1)!}
\end{aligned}
\end{equation}

\end{document}